\documentclass[a4paper]{article}
\usepackage[]{geometry}

\usepackage[utf8]{inputenc}

\usepackage{amsmath,amssymb}

\usepackage{cite}

\usepackage{nameref,hyperref}

\usepackage{microtype}
\usepackage{rotating}

\usepackage{calrsfs} %calligraphy \mathcal{L}
\DeclareMathAlphabet{\pazocal}{OMS}{zplm}{m}{n} %\pazocal{L}}

\makeatletter
\renewcommand{\@biblabel}[1]{\quad#1.}
\makeatother

\date{}

\begin{document}
\vspace*{0.35in}

% Please capitalize all terms in the title except conjunctions, prepositions, and articles.
\begin{flushleft}
{\Large
\textbf\newline{A three-threshold learning rule approaches the maximal capacity of recurrent neural networks}
}
\newline
\\
Alireza Alemi\textsuperscript{1,2,*},
Carlo Baldassi\textsuperscript{1,2},
Nicolas Brunel\textsuperscript{3},
Riccardo Zecchina\textsuperscript{1,2}
\\
\bigskip
\bf{1} Human Genetics Foundation (HuGeF), Turin, Italy
\\
\bf{2} DISAT, Politecnico di Torino, Turin, Italy
\\
\bf{3} Departments of Statistics and Neurobiology, University of Chicago, USA
\bigskip

% Insert additional author notes using the symbols described below. Insert symbol callouts after author names as necessary.
 
% Remove or comment out the author notes below if they aren't used.

% Use the asterisk to denote corresponding authorship and provide email address in note below.
* alemi@polito.it

\end{flushleft}
\section*{Abstract}
Understanding the theoretical foundations of how memories are encoded
and retrieved in neural populations is a central challenge in
neuroscience. A popular theoretical scenario for modeling memory
function is the attractor neural network scenario, whose prototype is
the Hopfield model. The model simplicity and the locality of the
synaptic update rules come at the cost of a poor storage capacity,
compared with the capacity achieved with perceptron learning
algorithms.  Here, by transforming the perceptron learning rule, we
present an  online learning rule for a recurrent neural network that
achieves near-maximal storage capacity without an explicit supervisory
error signal, relying only upon locally accessible information. The
fully-connected network consists of excitatory binary neurons with
plastic recurrent connections and non-plastic inhibitory feedback
stabilizing the network dynamics; the memory patterns to be memorized
are presented  online  as strong afferent currents, producing a bimodal
distribution for the neuron synaptic inputs. Synapses corresponding to
active inputs are modified as a function of the value of the local
fields with respect to three thresholds.  Above the highest threshold,
and below the lowest threshold, no plasticity occurs. In between these
two thresholds, potentiation/depression occurs when the local field is
above/below an intermediate threshold.  We simulated and analyzed a
network of binary neurons implementing this rule and measured its
storage capacity for different sizes of the basins of attraction. The
storage capacity obtained through numerical simulations is shown to be
close to the value predicted by analytical calculations. We also
measured the dependence of capacity on the strength of external
inputs. Finally, we quantified the statistics of the resulting
synaptic connectivity matrix, and found that both the fraction of zero
weight synapses and the degree of symmetry of the weight matrix
increase with the number of stored patterns.

\section*{Author Summary}
Recurrent neural networks have been shown to be able to store memory
patterns as fixed point attractors of the dynamics of the network. The
prototypical learning rule for storing memories in attractor neural
networks is Hebbian learning, which can store up to $0.138N$ uncorrelated
patterns in a recurrent network of $N$ neurons. This is very far from the
maximal capacity $2N$, which can be achieved by supervised rules, e.g.~by the
perceptron learning rule. However, these rules are problematic for neurons in
the neocortex or the hippocampus, since they rely on the computation of a
supervisory error signal for each neuron of the network. We show here that the
total synaptic input received by a neuron during the presentation of a
sufficiently strong stimulus contains implicit information about the error,
which can be extracted by setting three thresholds on the total input, defining
depression and potentiation regions. The resulting learning rule implements
basic biological constraints, and our simulations show that a network
implementing it gets very close to the maximal capacity, both in the dense and
sparse regimes, across all values of storage robustness. The rule predicts that
when the total synaptic inputs goes beyond a threshold, no potentiation should
occur.

%\linenumbers

\section*{Introduction}

One of the fundamental challenges in neuroscience is to understand how
we store and retrieve memories for a long period of time. Such
long-term memory is fundamental for a variety of our cognitive
functions. A popular theoretical framework for storing and retrieving
memories in recurrent neural networks is the attractor network model
framework \cite{hopfield82,amit89,hertz91}. Attractors, i.e.~stable
states of the dynamics of a recurrent network, are set by modification
of synaptic efficacies in a recurrent network. Synaptic plasticity
rules specify how the efficacy of a synapse is affected by pre- and
post-synaptic neural activity. In particular, Hebbian synaptic
plasticity rules lead to long-term potentiation (LTP) for correlated
pre- and post-synaptic activities, and long-term depression (LTD) for
anticorrelated activities. These learning rules build
excitatory feedback loops in the synaptic connectivity, resulting in
the emergence of attractors that are correlated with the patterns of
activity that were imposed  on the network through external
inputs. Once a set of patterns become attractors of a network (in
other words when the network ``learns'' the patterns), upon a brief
initial activation of a subpopulation of neurons, the network state
evolves towards the learned stable state (the network ``retrieves'' a
past stored memory), and remains in that state after removal of the
external inputs (and hence maintains the information in short-term
memory). The set of initial network states leading to a memorized
state is called the \textit{basin of attraction}, whose size
determines how robust a memory is. The attractor neural network
scenario was originally explored in networks of binary neurons
\cite{hopfield82,amit89}, and then extended from the 90s to networks
of spiking neurons \cite{amit97,brunel01b,mongillo08,barak14}.

Experimental evidence in different areas of the brain, including
inferotemporal cortex
\cite{fuster81,miyashita88,miyashita88b,nakamura95} and prefrontal
cortex \cite{fuster71,funahashi89,romo99}, has provided support for
the attractor neural network framework, using electrophysiological
recordings in awake monkeys performing delayed response tasks. In such
experiments, the monkey has to maintain information in short-term
(working) memory in a `delay period' to be able to perform the
task. Consistent with the attractor network scenario, some neurons
exhibit selective persistent activity during the delay period. This
persistent activity of ensembles of cortical neurons has thus been
hypothesized to form the basis of the working memory of stimuli shown
in these tasks.

One of the most studied properties of attractor neural network as a
model of memory is its storage capacity, i.e. how many random patterns
can be learned in a recurrent network of $N$ neurons in the large $N$
limit. Storage capacity depends both on the network architecture and
on the synaptic learning rule. In many models, the storage capacity
scales with $N$. In particular, the Hopfield network \cite{hopfield82}
that uses a Hebbian learning rule has a storage capacity of $0.138N$
in the limit of $N\rightarrow\infty$ \cite{amit85}. Later studies
showed how the capacity depends on the connection probability in a
randomly connected network \cite{sompolinsky86,derrida87} and on the
coding level (fraction of active neurons in a pattern)
\cite{tsodyks88,buhmann89}. A natural question is, what is the maximal
capacity of a given network architecture, over all possible learning
rules? This question was answered by Elizabeth Gardner, who showed
that the capacity of fully connected networks of binary neurons with
dense patterns scales as $2N$ \cite{gardner88}, a storage capacity
which is much larger than the one of the Hopfield model. The next
question is what learning rules are able to saturate the Gardner
bound? A simple learning rule that is guaranteed to achieve this bound is the
perceptron learning rule (PLR) \cite{rosenblatt62} applied to each neuron
independently. However, unlike the rule used in the Hopfield model, the
perceptron learning rule is a supervised rule that needs an explicit ``error
signal'' in order to achieve the Gardner bound. While such an error signal
might be available in the cerebellum \cite{marr69,albus71,ito82}, it is unclear
how error signals targeting individual neurons might be implemented in cortical
excitatory synapses. Therefore, it remains unclear whether and how networks
with realistic learning rules might approach the Gardner bound.

The goal of the present paper is to propose a learning rule whose
capacity approaches the maximal capacity of recurrent neural networks
by transforming the original perceptron learning rule such that the
new rule does not explicitly use an error signal. The perceptron
learning rule modifies the synaptic weights by comparing the desired
output with the actual output to obtain an error signal, subsequently
changing the weights in the opposite direction of the error signal. We
argue that the total synaptic inputs (`local fields') received by a
neuron during the presentation of a stimulus contain some information
about the current error (i.e. whether the neuron will end up in the
right state after the stimulus is removed). We use this insight to
build a field dependent learning rule that contains three thresholds
separating no plasticity, LTP and LTD regions. This rule implements
basic biological constraints: (a) it uses only information local to
the synapse; (b) the new patterns can be learned incrementally, i.e. it is an online rule; (c) it
does not need an explicit error signal; (d) synapses obey Dale's
principle, i.e. excitatory synapses are not allowed to have negative
weights.  We studied the capacity and the size of the basins of
attraction for a binary recurrent neural network in which excitatory
synapses are endowed with this rule, while a global inhibition term controls
the global activity level. We investigated how the strength of external fields
and the presence of correlations in the inputs affect the memory capacity.
Finally, we investigated the statistical properties of the connectivity matrix
(distribution of synaptic weights, degree of symmetry).

%------------------------------------------------------
% RESULTS
%
% Results and Discussion can be combined.
\section*{Results}

\subsection*{The network}

\begin{figure}
%\internallinenumbers
\begin{center}
%\begin{tabular}{l}

\includegraphics[scale=1]{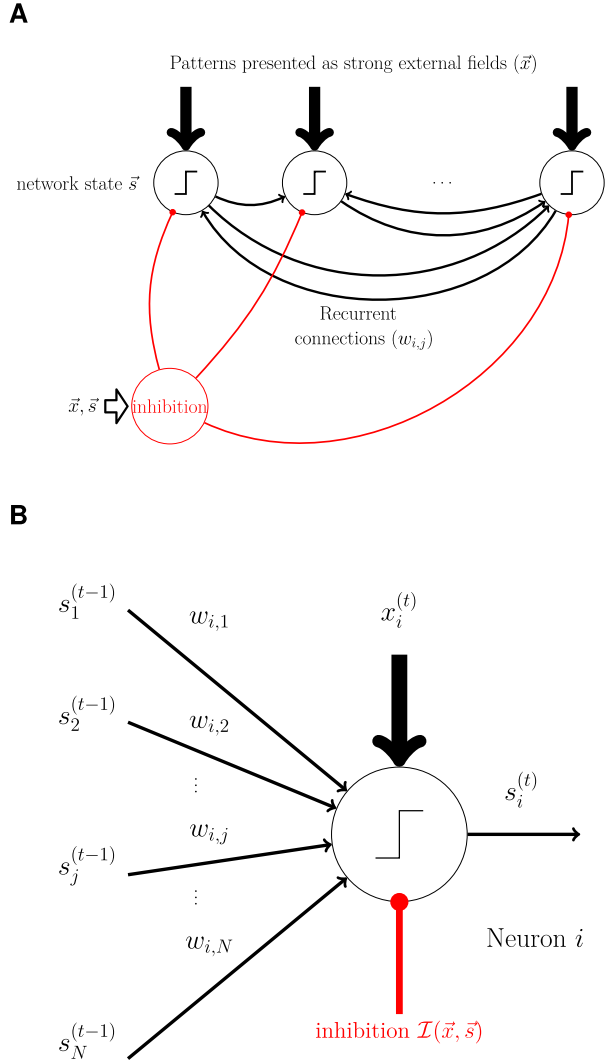}

\end{center}
\caption{
\textbf{A sketch of the network and the neuron model.}
  \textbf{A.} Structure of the network. The fully-connected network
  consists of $N$ binary $(s_i\in\{0,1\})$ neurons and an aggregated
  inhibitory unit. The global inhibition is a function of the state of
  the network and the external fields,
  i.e. $\pazocal{I}(\vec{x},\vec{s})$. A memory pattern $\vec{\xi}$ is
  encoded as strong external fields, i.e. $\vec{x}=X\vec{\xi}$ and
  presented to the network during the learning phase.  \textbf{B.}
  Each neuron receives excitatory recurrent inputs (thin black arrows)
  from the other neurons, a global inhibitory input (red connections),
  and a strong binary external field ($x_i\in\{0,X\}$; thick black
  arrows). All these inputs are summed to obtain the total field,
  which is then compared to a neuronal threshold $\theta$; the output
  of the neuron is a step function of the result.}
\label{fig:network}
\end{figure}
We simulated a network of $N$ binary (McCulloch-Pitts) neurons, fully-connected
with excitatory synapses (Fig~\ref{fig:network}A). All the neurons feed a
population of inhibitory neurons which is modeled as a single aggregated
inhibitory unit. This state-dependent global inhibition projects back onto all
the neurons, stabilizing the network and controlling its activity level.  At
each time step, the activity (or the state) of neuron $i$ ($i=1\ldots N$) is
described by a binary variable $s_{i}\in\{0,1\}$. The state is a step function
of the \textit{local field} $v_{i}$ of the neuron:
\begin{equation}
	s_{i} = \Theta\left(v_{i}-\theta\right),\label{eq:theta}
\end{equation}
where $\Theta$ is the Heaviside function ($\Theta\left(x\right)=1$ if $x>0$ and
$0$ otherwise) and $\theta$ is a neuronal threshold. The local field $v_{i}$
represents the overall input received by the neuron from its excitatory and
inhibitory connections (Fig~\ref{fig:network}B). The excitatory connections
are of two kinds: recurrent connections from within the excitatory population,
and external inputs.

The recurrent excitatory connections are mediated by synaptic weights, denoted
by a matrix $W$ whose elements $w_{ij}$ (the weight of the synapse from neuron
$j$ to $i$) are continuous non-negative variables ($w_{ij} \in [ 0,\infty)$;
$w_{ii}=0$). In the following, and in all our simulations, we assume that the
weights are initialized randomly before the training takes place (see
Materials and Methods).

Therefore, in the absence of external inputs, the local field of each
neuron $i$ is given by:
\begin{equation}
v_{i}=\sum_{j=1}^N{w_{ij} s_{j}} - \pazocal{I}_0\left(\vec{s}\right),\label{eq:vi0}
\end{equation}
where $\pazocal{I}_0\left(\vec{s}\right)$ represents the inhibitory input.

For the sake of simplicity, we simulated a synchronous update process,
in which the activity of each neuron $s_i$ is computed from the local
field $v_i$ at the previous time step, and all updates happen in
parallel.

The network was designed so that, in absence of external input and
prior to the training process, it should spontaneously stabilize
itself to some fixed overall average activity level $f$ (fraction of
active neurons, or sparseness), regardless of the initial
conditions. In particular, we aimed at avoiding trivial attractors
(the all-off and all-on states). To this end, we model the inhibitory
feedback (in absence of external inputs) as a linear function of the
overall excitatory activity:
\begin{equation}
\pazocal{I}_0(\vec{s}) = H_0 + \lambda\left(\sum_{i=1}^N{s_{i}} - f N\right).\label{eq:I0}
\end{equation}
The parameters $H_0$ and $\lambda$ can be understood as follows: $H_0$
is the average inhibitory activity when the excitatory network has the
desired activity level $f$, i.e. when $\sum_{i=1}^N{s_{i}}=f N$;
$\lambda$ measures the strength of the inhibitory feedback onto the
excitatory network. This expression can be interpreted as a
first-order approximation of the inhibitory activity as a function of
the excitatory activity around some reference value $f N$, which is
reasonable under the assumption that the deviations from $f N$ are
small enough. Indeed, by properly setting these two parameters in
relation to the other network parameters (such as $\theta$ and the
average connection strength) it is possible to achieve the desired
goal of a self-stabilizing network.

In the training process, the network is presented a set of $p$
patterns in the form of strong external inputs, representing the
memories which need to be stored. We denote the patterns as
$\{\vec{\xi}^{\mu}\}$ (where $\mu=1...p$ and
$\xi_{i}^{\mu}\in\{0,1\}$), and assume that each entry $\xi_{i}^{\mu}$
is drawn randomly and independently. For simplicity, the coding level
$f$ for the patterns was set equal to the spontaneous activity level
of the network, i.e.~ $\xi_{i}^{\mu}=1$ with probability $f$, 0
otherwise. During the presentation of a pattern $\mu$, each neuron $i$
receives an external binary input $x_{i} = X\xi_{i}^{\mu}$, where $X$
denotes the strength of the external inputs, which we parameterized as
$X=\gamma \sqrt{N}$. In addition, the external input also affects the
inhibitory part of the network, eliciting a response which indirectly
downregulates the excitatory neurons. We model this effect as an
additional term $H_1$ in the expression for the inhibitory term
(Eq.~\ref{eq:I0}), which therefore becomes:
\begin{equation}
	\pazocal{I}(\vec{x},\vec{s})=H_0 + H_1\frac{\sum_{i=1}^N{x_i}}{fNX}  + \lambda(\sum_{i=1}^N{s_{i}} - fN),\label{eq:I}
\end{equation}
The general expression for the local field $v_i$ then reads:
\begin{equation}
	v_{i}=\sum_{j=1}^N{w_{ij} s_{j}} + x_{i} - \pazocal{I}(\vec{x},\vec{s}).\label{eq:vi}
\end{equation}
In the absence of external fields, $x_i=0$ for all $i$, and thus
Eqs.~\ref{eq:I} and~\ref{eq:vi} reduce to the previous expressions
Eqs.~\ref{eq:I0} and~\ref{eq:vi0}.

The goal of the learning process is to find values of $w_{ij}$'s such
that the patterns $\{\vec{\xi}^{\mu}\}$ become attractors of the
network dynamics. Qualitatively, this means that, if the training
process is successful, then whenever the network state gets
sufficiently close to one of the stored patterns, i.e.~whenever the
Hamming distance $d = \sum_{i=1}^{N}\left|\xi_{i}^{\mu} -
  s_{i}\right|$ between the current network state and a pattern $\mu$
is sufficiently small, the network dynamics in the absence of external
inputs should drive the network state towards a fixed point equal to
the pattern itself (or very close to it). The general underlying idea
is that, after a pattern is successfully learned, some brief external
input which initializes the network close to the learned state would
be sufficient for the network to recognize and retrieve the
pattern. The maximum value of $d$ for which this property holds is
then called the basin of attraction size (or just basin size hereafter
for simplicity); indeed, there is generally a trade-off between the
number of patterns which can be stored according to this criterion and
the size of their basin of attraction.

More precisely, the requirement that a pattern $\vec{\xi}^{\mu}$ is a fixed point of the network dynamics in the absence of external fields can be reduced to a condition for each neuron $i$ (cfr.~Eqs.~\ref{eq:I} and~\ref{eq:vi}):
\begin{equation}
	\forall{i}:\ \Theta\left(\sum_{j=1}^{N}{w_{ij} \xi_{j}^{\mu}} - \pazocal{I}\left(\vec{0},\vec{\xi}^{\mu}\right) - \theta\right) = \xi_{i}^{\mu}.\label{eq:fixedpoint}
\end{equation}
This condition only guarantees that, if the network is initialized
into a state $\vec{s}=\vec{\xi^{\mu}}$, then it will not spontaneously
change its state, i.e.~it implements a zero-size basin of
attraction. A simple way to enlarge the basin size is to make the
requirement in Eq.~\ref{eq:fixedpoint} more stringent, by enforcing a
more stringent constraint for local fields:
\begin{eqnarray}
	\forall{i}:\ \begin{cases}
		\sum_{j=1}^{N}{w_{ij} \xi_{j}^{\mu}} - \pazocal{I}\left(\vec{0},\vec{\xi^{\mu}}\right) > \theta + f\sqrt{N}\epsilon & \text{if } \xi_i^\mu = 1\\
		\sum_{j=1}^{N}{w_{ij} \xi_{j}^{\mu}} - \pazocal{I}\left(\vec{0},\vec{\xi^{\mu}}\right) < \theta - f\sqrt{N}\epsilon & \text{if } \xi_i^\mu = 0,
	\end{cases}\label{eq:fixedpointeps}
\end{eqnarray}
where $\epsilon\ge0$ is a robustness parameter. When $\epsilon=0$, we
recover the previous zero-basin-size scenario; increasing $\epsilon$
we make the neurons' response more robust towards noise in their
inputs, and thus we enlarge the basin of attraction of the stored
patterns (but then fewer patterns can be stored, as noted above).

\subsection*{The three-threshold learning rule (3TLR)}

In the training phase, the network is presented with patterns as
strong external fields $x_{i}$. Patterns are presented sequentially in
random order. For each pattern $\mu$, we simulated the following
scheme:

{\bf Step 1}: The pattern is presented (i.e.~the external inputs
$x_{i}$ are set to $X \xi_i^\mu$). A single step of synchronous
updating is performed (Eqs.~\ref{eq:theta}, \ref{eq:I}
and~\ref{eq:vi}). If the external inputs are strong enough,
i.e.~$\gamma$ is large enough, this updating sets the network in a
state corresponding to the presented pattern.

{\bf Step 2}: Learning occurs. Each neuron $i$ may update its synaptic
weights depending on 1) their current value $w_{ij}^t$, 2) the state
of the pre-synaptic neurons, and 3) the value of the local field
$v_i$. Therefore, all the information required is locally accessible,
and no explicit error signals are used. The new synaptic weights
$w_{ij}^{t+1}$ are set to:
\begin{equation}
	w_{ij}^{t+1}=
	\begin{cases}
		w_{ij}^t - \eta s_j, & \text{if } \theta_0 < v_i < \theta \\

		w_{ij}^t + \eta s_j, & \text{if } \theta < v_i < \theta_1 \\

		w_{ij}^t ,&  \text{otherwise, }
	\end{cases}\label{eq:update}
\end{equation}
where $\eta$ is the learning rate, and $\theta_0$ and $\theta_1$ are
two auxiliary learning thresholds set as
\begin{eqnarray}
	\theta_0 & = & \theta - \left(\gamma + \epsilon\right)f\sqrt{N}\label{eq:theta0} \\
	\theta_1 & = & \theta + \left(\gamma + \epsilon\right)f\sqrt{N}.\label{eq:theta1}
\end{eqnarray}

We refer to this update scheme as ``three-threshold learning rule'' (3TRL).
After some number of presentations, we checked whether the patterns
are learned by presenting a noisy version of these patterns, and
checking whether the patterns (or network states which are very close
to the patterns) are fixed points of the network dynamics.

%fig:rule
\begin{figure}
%\internallinenumbers
\begin{center}
\includegraphics[scale=1]{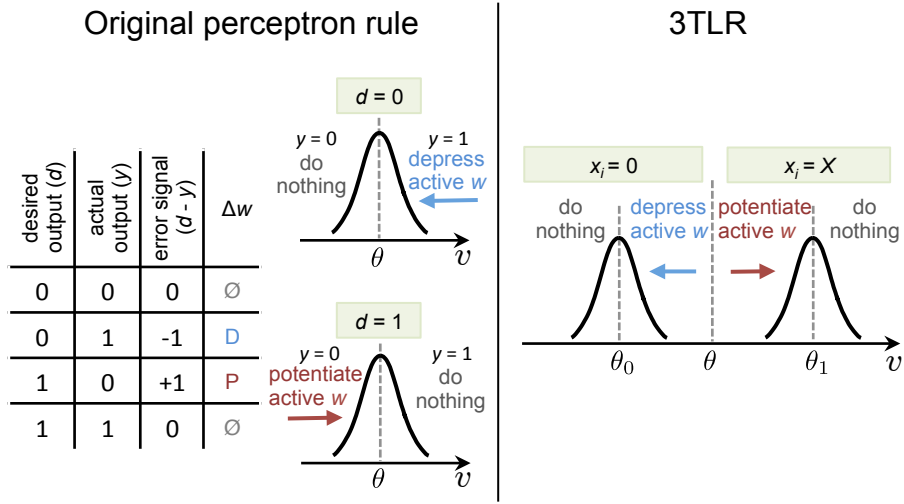}
\end{center}
\caption{\textbf{The three-threshold learning rule (3TLR), and its relationship with the
  standard perceptron learning rule (PLR).} The perceptron learning
  rule modifies the synaptic weights by comparing the desired output
  with the actual output to obtain an error signal, subsequently
  changing the weights in the opposite direction of the error signal
  (see the table in the left panel). For a pattern which is
  uncorrelated with the current synaptic weights, the distribution is
  Gaussian (in the limit of large $N$), due to the central limit
  theorem. $H_0$ is set such that, on average, a fraction $f$ of the
  local fields are above the neuronal threshold $\theta$; in the case
  of $f=0.5$, this means that the Gaussian is centered on $\theta$
  (left panel). In our model (Fig~\ref{fig:network}B), the desired
  output is given as a strong external input, whose distribution
  across the population is bimodal (with two delta
    functions on $x_i=0$ and $x_i=X$); therefore, the distribution of
  the local fields during stimulus presentation becomes bimodal as
  well (right panel). The left and right bumps of this distribution
  correspond to cases where the desired outputs are zero and one,
  respectively. Note that, since the external input also elicits an
  inhibitory response, the neurons in the network which are not
  directly affected by the external input (i.e.~those with desired
  output equal to zero) are effectively hyperpolarized. If $X$ is
  sufficiently large, the two distributions do not overlap, and the
  four cases of the PLR can be mapped to the four regions determined
  from the three thresholds, indicated by vertical dashed lines (see
  text).}
\label{fig:rule}
\end{figure}

 %fig:potentials
\begin{figure}[!ht]
%\internallinenumbers
\begin{center}
\includegraphics[width=0.8\linewidth]{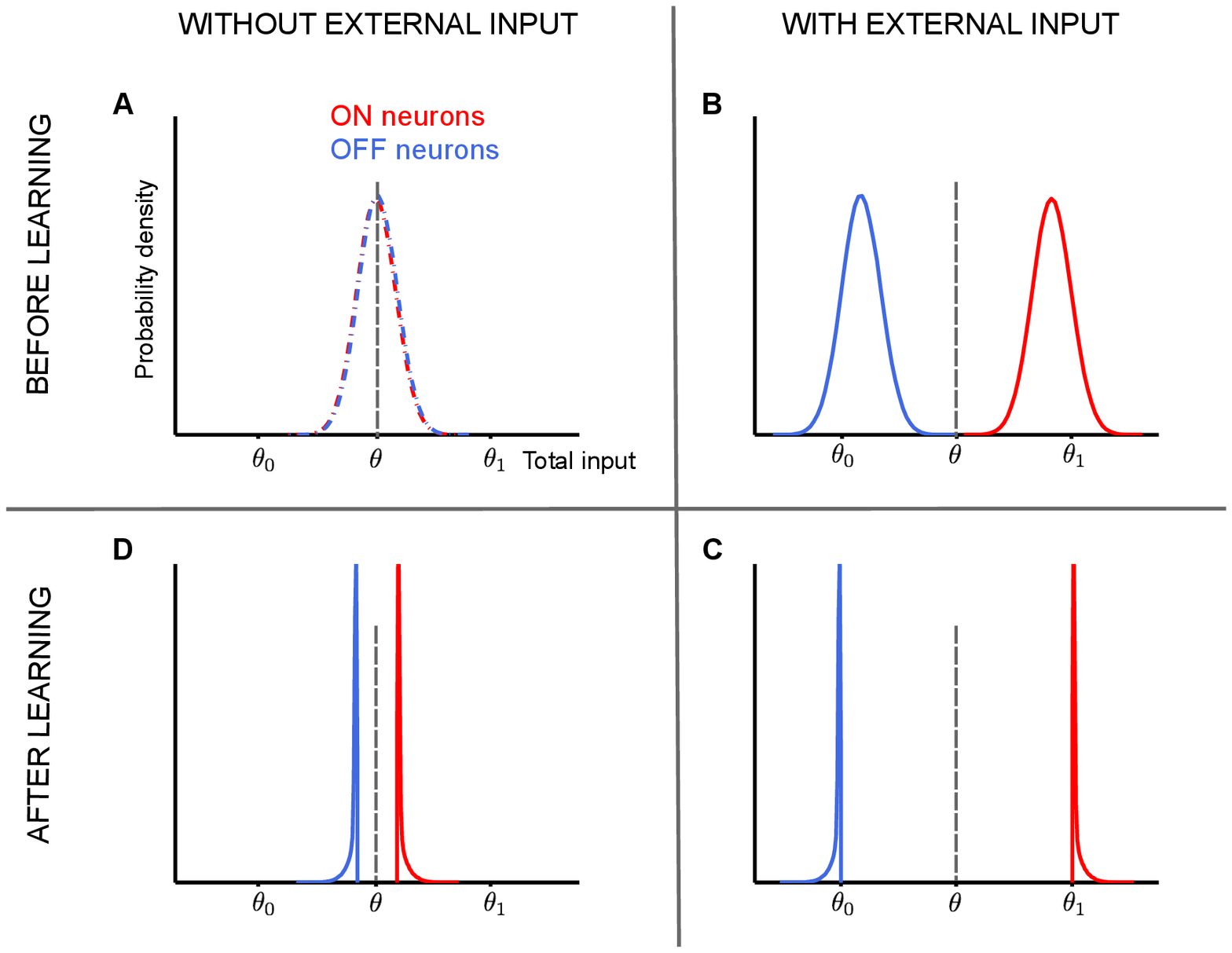}
\end{center}
\caption{\textbf{Distribution of local fields before and after learning
  for $f=0.5$ and non-zero robustness.} \textbf{A.} Before learning
  begins, the distribution of local field of neurons is a Gaussian
  distribution (due to central limit theorem) centered around neuronal
  threshold $\theta$ both for neurons with the desired output zero
  (OFF neurons) and with the desired output one (ON neurons). The goal
  is to have the local field distribution of ON neurons (red curve) to
  be above the threshold $\theta$, and that of OFF neurons to be below
  $\theta$. \textbf{B.} Once any of the to-be-stored patterns are
  presented as strong external fields, right before the learning
  process starts, the local field distribution of the OFF neuron
  shifts toward the left-side centered around $\theta_0 + f \epsilon
  \sqrt{N}$, whereas the distribution of the ON neurons moves toward
  the right-side, centered around $\theta_1 - f \epsilon \sqrt{N}$,
  with a negligible overlap between the two curves if the external
  field is strong enough. Thanks to the strong external fields and
  global inhibition, the local fields of the ON and OFF neurons are
  well separated. \textbf{C.} Due to the learning process, the local
  fields within the depression region [i.e. $(\theta_0,\theta)$] get
  pushed to the left-side, below $\theta_0$, whereas those within the
  potentiation region get pushed further to the right-side, above
  $\theta_1$. If the learning process is successful, it will result in
  a region $(\theta_0,\theta_1)$ which no longer contain local fields,
  with two sharp peaks on $\theta_0$ and $\theta_1$. \textbf{D.} After
  successful learning, once the external fields are removed, the blue
  and red curves come closer, with a gap equal to $2 f \epsilon
  \sqrt{N}$. The larger the robustness parameter $\epsilon$, the more
  the gap between the left- and right-side of the distribution. Notice
  that now the red curve is fully above $\theta$ which means those
  neurons remain stably ON, while the the blue curve is fully below
  $\theta$, which means those neurons are stably OFF. Therefore the
  corresponding pattern is successfully stored by the network.}

\label{fig:potentials}
\end{figure}

When $N\gg 1$, $\gamma$ is large enough, and $H_1=f X$, the update rule
described by Eq.~\ref{eq:update} is essentially equivalent to the
perceptron learning rule for the task described in
Eq.~\ref{eq:fixedpointeps}. This can be shown as follows (see also
Fig~\ref{fig:rule} for a graphical representation of the case
$f=0.5$ and $\epsilon=0$): when a stimulus is presented, the
population of neurons is divided in two groups, one for which $x_i=0$
and one for which $x_i=X$. The net effect of the stimulus presentation
on the local field has to take into account the indirect effect
through the inhibitory part of the network (see Eq.~\ref{eq:I}), and
thus is equal to $-f X$ for the $x_i=0$ population and to
$\left(1-f\right)X$ for the $x_i=X$ population. Before learning, the
distribution of the local fields across the excitatory population, in
the limit $N\to\infty$, is a Gaussian whose standard deviation is
proportional to $\sqrt{N}$, due to the central limit theorem;
moreover, the parameter $H_0$ is set so that the average activity
level of the network is $f$, which means that the center of the
Gaussian will be within a distance of order $\sqrt{N}$ from the
neuronal threshold $\theta$ (this also applies if we use
different values for the spontaneous activity level and the pattern
activity level).  Therefore, if $X=\gamma\sqrt{N}$ is large enough,
the state of the network during stimulus presentation will be
effectively clamped to the desired output, i.e.~$s_i=\xi_i^\mu$ for
all $i$. This fact has two consequences: 1) the local field potential
can be used to detect the desired output by just comparing it to the
threshold, and 2) each neuron $i$ will receive, as its recurrent
inputs $\left\{s_j\right\}_{j\ne i}$, the rest of the pattern
$\left\{\xi_j^\mu\right\}_{j\ne i}$.  Furthermore, due to the choice
of the secondary thresholds $\theta_0$ and $\theta_1$ in
Eqs.~\ref{eq:theta0} and~\ref{eq:theta1}, the difference between the
local field and $\theta_0$ (or $\theta_1$) during stimulus
presentation for the $x_i=0$ population (or $x_i=X$, respectively) is
equal to the difference between the local field and $\theta - f
\sqrt{N}\epsilon$ (or $\theta + f \sqrt{N}\epsilon$, respectively) in
the absence of external stimuli, provided the recurrent inputs are the
same. Therefore, the value of the local field $v_i$ during stimulus
presentation in relation to the three thresholds $\theta$, $\theta_0$
and $\theta_1$ is sufficient to determine whether an error is made
with respect to the constraints of Eq.~\ref{eq:fixedpointeps}, and
which kind of error is made. Following these observations, it is
straightforward to map the standard perceptron learning rule on the 4
different cases which may occur (see Fig~\ref{fig:rule}), resulting
in Eq.~\ref{eq:update}.

In Fig~\ref{fig:potentials} we demonstrate the effect of the
learning rule on the distribution of the local field potentials as
measured from a simulation (with $f=0.5$ and $\epsilon=1.2$): the
initial distribution of the local fields of the neurons, before the
learning process takes place and in the absence of external fields, is
well described by a Gaussian distribution centered on the neuronal
threshold $\theta$ (see Fig~\ref{fig:potentials}A) with a standard
deviation which scales as $\sqrt{N}$. During a pattern presentation,
the resulting distribution becomes a bimodal one; before learning
takes place, the distribution is given by the sum of two Gaussians of
equal width, centered around $\theta_0+f\sqrt{N} \epsilon $ and
$\theta_1-f \sqrt{N} \epsilon$ (Fig~\ref{fig:potentials}B). The
left Gaussian corresponds to the cases where $x_i=0$ and the right one
to the cases where $x_i=X$. Having applied the learning rule, we
observe that the depression region (i.e.~the interval
$(\theta_0,\theta)$) and the potentiation region
(i.e.~$(\theta,\theta_1)$) gets depleted
(Fig~\ref{fig:potentials}C). In the testing phase, when the
external inputs are absent, the left and right parts of the
distribution come closer, such that the distance between the two peaks
is equal to at least $2\epsilon f \sqrt{N}$
(Fig~\ref{fig:potentials}D). This margin between the local fields
of the ON and OFF neurons makes the attractors more robust.

\subsection*{Storage capacity}

Since our proposed learning rule is able to mimic (or approximate,
depending on the parameters) the perceptron learning rule, which is
known to be able to solve the task posed by Eq.~\ref{eq:fixedpointeps}
whenever a solution exists, we expect that a network implementing such
rule can get close to maximal capacity in terms of the number of
memories which it can store at a given robustness level.  The storage
capacity, denoted by $\alpha=p/N$, is measured as a ratio of the
maximum number of patterns $p$ which can successfully be stored to the
number of neurons $N$, in the limit of large $N$.  As mentioned above,
it is a function of the basin size.

 %fig:capacitybasin
\begin{figure}[!ht]
%\internallinenumbers
%\begin{picture}(0,0)
%\put(50,-15){{\fontfamily{phv}\selectfont \Large{\textbf{A}} }}\end{picture}
\begin{center}
\includegraphics[scale=1]{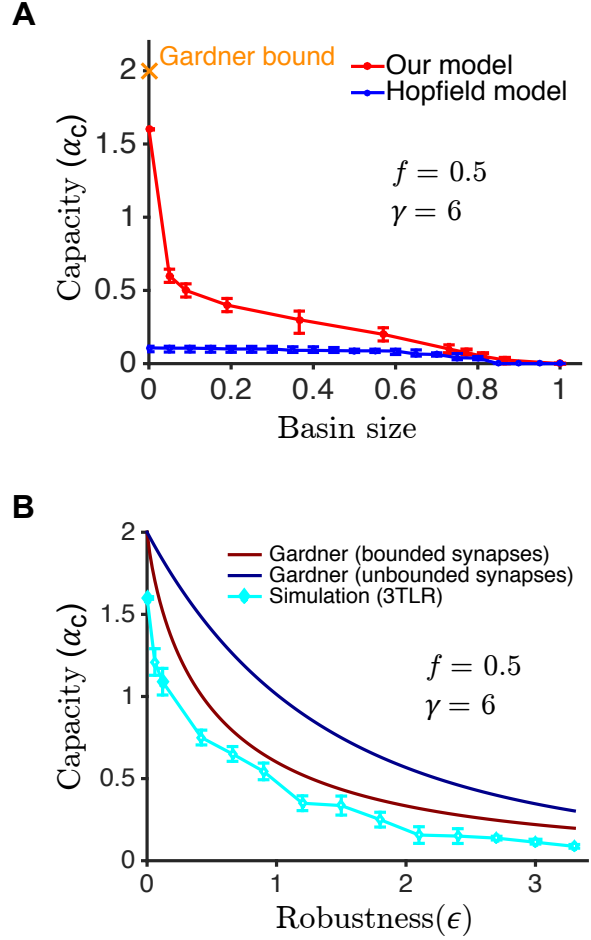}
%\includegraphics[width=.5\linewidth]{figs/AlphaVsBasin.png}
%\end{center}
%\begin{picture}(0,0)\put(50,-15){{\fontfamily{phv}\selectfont \Large{\textbf{B}} }}\end{picture}
%\begin{center}
%\includegraphics[width=.5\linewidth]{figs/alpha_vs_eps_dense.png}
\end{center}
\caption{\textbf{Critical capacity as a function of the basin size
  and the robustness parameter.}  \textbf{A.} The red plot shows the
  critical capacity as a function of the size of the basins of
  attraction ($N=1001$ neurons in the dense regime $f=0.5$) when the
  strength of the external field is large ($\gamma=6$) such that the
  ON and OFF neuronal populations are well separated. The points
  indicate $0.5$ probability of successful storage at a given basin
  size, optimized over the robustness parameter $\epsilon$ . The error
  bars show the $[0.95,0.05]$ probability interval for successful
  storage. The blue plot shows the performance of the Hopfield model
  with $N=1001$ neurons. The maximal capacity at zero basin size (the
  Gardner bound) is equal to $2$.  \textbf{B.} To compare the result
  of simulation of our model with the analytical results, we plotted
  the critical capacity as a function of the robustness parameter
  $\epsilon$. The dark red curve is the critical capacity versus
  $\epsilon$ for our model obtained form analytical calculations (see
  Materials and Methods), the cyan line shows the result of
  simulations of our model, and the dark blue shows the Gardner bound
  for a network with no constraints on synaptic weights. The
  difference between the two theoretical curves is due to the
  constraints on the weights in our network.}
\label{fig:capacitybasin}
\end{figure}

We used the following definition for the basin size: a set of $p$
patterns is said to be successfully stored at a size $b$ if, for each
pattern, the retrieval rate when starting from a state in which a
fraction $b$ of the pattern was randomized is at least $90\%$.  The
retrieval rate is measured by the probability that the network
dynamics is able to bring the network state to an attractor within
$1\%$ distance from the pattern, in at most $30$ steps.  The distance
between the state of the network and a pattern $\mu$ is measured by
the normalized Hamming distance
$\frac{1}{N}\sum_{i=1}^N\left|s_i-\xi_i^\mu\right|$.  Therefore, at
coding level $f=0.5$, reaching a basin size $b$ means that the network
can successfully recover patterns starting from a state at distance
$b/2$.

Fig~\ref{fig:capacitybasin}A shows the maximal capacity as a
function of the basin size for a simulated network of $N=1001$
neurons. We simulated many pairs of ($\alpha,\epsilon$) with different
random seeds, obtaining a probability of success for each pair.  The
red line shows the points for which the probability of successful
storage is $0.5$, and the error bars span $0.95$ to $0.05$ success
probability. The capacity was optimized over the robustness parameter
$\epsilon$. The maximal capacity (the Gardner bound) in the limit of
$N\rightarrow \infty$ at the zero basin size is $\alpha_c=2$ for our
model (see Materials and Methods for the calculation), as for a
network with unconstrained synaptic weights \cite{gardner88}.  In
Fig~\ref{fig:capacitybasin}A, we also compare our network with the
Hopfield model.  Our network stores close to the maximal capacity at
zero basin size, at least eleven times more than the Hopfield
model. Across the range of basin sizes, 3TLR achieves
more than twice the capacity that can be achieved with the Hopfield
model. 

The enlargement of the basin of attraction was achieved by increasing
the robustness parameter $\epsilon$.  We computed the maximal
theoretical capacity as a function of $\epsilon$ at $N\to\infty$ (see
Materials and Methods) and compared it to our simulations, and to the
maximal theoretical capacity of the Hopfield network.  The results are
shown in Fig~\ref{fig:capacitybasin}B.  For any given value of
$\epsilon$, the cyan curve shows the maximum $\alpha$ for which the
success ratio with our network was at least $0.5$ across different
runs.  The difference between the theory and the experiments in our
model can be ascribed to several factors: the finite size of the
network; the choice of the finite learning rate $\eta$, and the fact
that we imposed a hard limit on the number of pattern presentations
(see number of iterations in Table~\ref{tab:param}), while the
perceptron rule for excitatory synaptic connectivity is only
guaranteed to be optimal in the limit of $\eta\to0$, with a number of
presentations inversely proportional to $\eta$ \cite{clopath12}.
Note that the correspondence between the PLR and the 3TLR is only
  perfect in the large $\gamma$ limit, and is only approximate
  otherwise, as can be shown by comparing explicitly the synaptic
  matrices obtained by both algorithms on the same set of patterns
  (see Materials and Methods.)

 %fig:gamma
\begin{figure}[!ht]
%\internallinenumbers
\begin{center}
\includegraphics[width=0.6\linewidth]{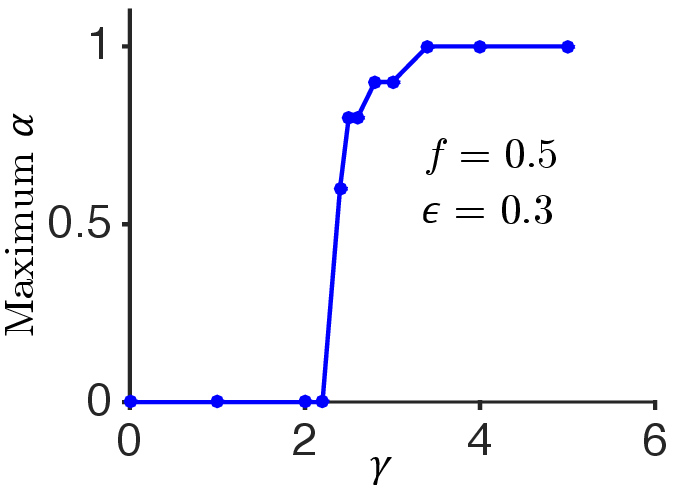}
\end{center}
\caption{\textbf{Dependence of the critical capacity on the strength of
  the external input.}  We varied the strength of the external field
  ($\gamma$) in order to quantify its effect on the learning
  process. The critical capacity is plotted as a function of $\gamma$
  at a fixed robustness $\epsilon=0.3$ in the dense regime
  $f=0.5$. The simulations show that there is a very sharp drop in the
 maximum $\alpha$ when $\gamma$  goes below $\approx 2.4$.  }
\label{fig:gamma}
\end{figure}

A crucial ingredient of the 3TLR is having a strong external
input which effectively acts as a supervisory signal. How strong do
the external fields need to be? How much does the capacity depend on this
strength? To answer these questions, we measured the maximum number of stored
patterns as a function of the parameter $\gamma$ which determines the strength
of external fields as $X=\gamma\sqrt{N}$.  This parameter, in fact, determines
how far the two Gaussian distributions of the local field are; as shown in
Fig~\ref{fig:rule}, the distance between the two peaks of the distribution is
$X$. For large enough $\gamma$, the overlap of these two distributions is
negligible and the capacity is maximal; but as we lower $\gamma$, the overlap
increases, causing the learning rule to make mistakes, i.e.~when it should
potentiate, it depresses the synapses and vice versa.  In our simulations with
$N=1001$ neurons in the dense regime $f=0.5$ at a fixed epsilon $\epsilon=0.3$,
we varied $\gamma$ and computed the maximum $\alpha$ that can be achieved with
a fixed number of iterations ($1000$). The capacity indeed gradually decreases
as $\gamma$ decreases, until it reaches a threshold, below which there is a
sharp drop of capacity (see Fig~\ref{fig:gamma}). With the above values for the parameters, this
transition occurs at $\gamma \approx 2.4$.

%fig:sparse
\begin{figure}[!ht]
%\internallinenumbers
\begin{center}
\includegraphics[width=0.5\linewidth]{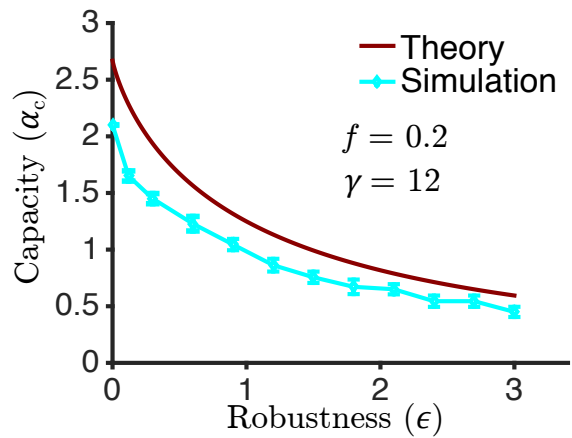}
\end{center}
\caption{\textbf{Capacity as a function of the robustness parameter
  $\epsilon$ at sparseness $f=0.2$.} The theoretical calculations is
  compared with the simulations for $f=0.2$. Note that the capacity in
  the sparse regime is higher than in the dense regime.}
\label{fig:sparse}
\end{figure}

The 3TLR can also be adapted to work in a sparser regime, at
a coding level lower than $0.5$.  However, the average activity level
of the network is determined by $H_0$, and their relationship also
involves the variance of the distribution of the synaptic weights when
$f\ne0.5$ (see Materials and Methods).  During the learning process,
the variance of the weights changes, which implies that the parameter
$H_0$ must adapt correspondingly.  In our simulations, this adaptation
was performed after each complete presentation of the whole pattern
set.  In practice, this additional self-stabilizing mechanism could
still be performed in an unsupervised fashion along with (or in
alternation with) the learning process.  Using this adjustment, we
simulated the network at $f=0.2$ and compared the results with the
theoretical calculations.  As shown in Fig~\ref{fig:sparse}, we can
achieve at least $70\%$ of the critical capacity across different values
of the robustness parameter $\epsilon$.

%fig:correlated
\begin{figure}[!ht]
%\internallinenumbers
\begin{center}
\includegraphics[width=0.5\linewidth]{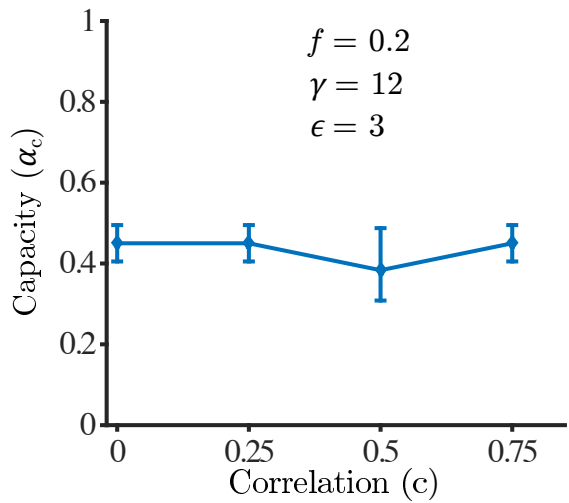}
\end{center}
\caption{\textbf{Capacity as a function of correlations in the
      input patterns, for $f=0.2$ at $\epsilon=3.0$.}
    Patterns are organized in categories, with a correlation $c$ with
    the prototype of the corresponding category (see text).}
\label{fig:correlated}
\end{figure}

We also investigated numerically the effect of correlations in the
input patterns. The PLR is able to learn correlated patterns as
long as a solution to the learning problem exists. As the 3TLR
approximates the PLR, we expect the 3TLR to be able to learn
correlated patterns as well. As a simple model of correlation, we
tested patterns organized in $L$ categories
\cite{parga86,brunel98b}. Each category was defined by a randomly
generated prototype. Prototypes were uncorrelated from category to
category. For each category, we then generated $p/L$ patterns
independently with a specified correlation coefficient $c$ with the
corresponding prototype. We show in Fig~\ref{fig:correlated} the
results of simulations with $L=5$, $f=0.2$ and $\epsilon=3$. The
figure shows that the learning rule reaches a capacity that is
essentially independent of $c$, in the range $0\le c \le 0.75$.

\subsection*{Statistical properties of the connectivity matrix}

\begin{figure}[!ht]
%\internallinenumbers
\begin{center}
\includegraphics[width=.9\linewidth]{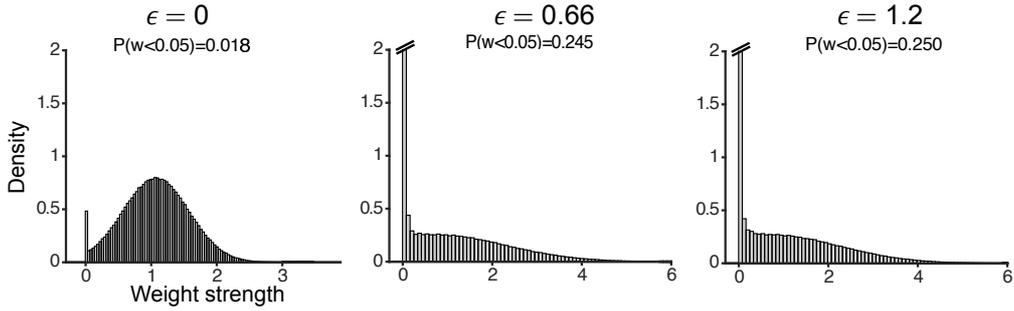}
\end{center}
\caption{\textbf{Synaptic weight distributions.} Comparing the distributions of
  the synaptic weights at critical capacity for three different values
  of robustness obtained from simulation. The distribution of weights
  approaches a Dirac-delta distribution at zero plus a truncated
  Gaussian. As the patterns become more robust, the center of the
  partial Gaussian shifts towards the left, and the number of silent
  synapses increases.}
\label{fig:weightdistribution}
\end{figure}

We next investigated the statistical properties of the connectivity
matrix after the learning process. Previous studies have shown that
the distribution of synaptic weights in perceptrons with excitatory
synapses becomes at maximal capacity a delta function at zero weight,
plus a truncated Gaussian for strictly positive weights
\cite{brunel04,brunel07,clopath12,clopath13}. Our model differs from
this setting because of the global inhibitory feedback. Despite this
difference, the distribution of weights in our network bear
similarities with the results obtained in these previous studies: the
distribution exhibits a peak at zero weight (`silent', or `potential'
synapses), while the distribution of strictly positive weights
resembles a truncated Gaussian. Finally, the fraction of silent
synapses increases with the robustness parameter (see
Fig~\ref{fig:weightdistribution}).

%fig:weightdistribution

\begin{figure}[!ht]
%\internallinenumbers
\begin{center}
\includegraphics[width=0.6\linewidth]{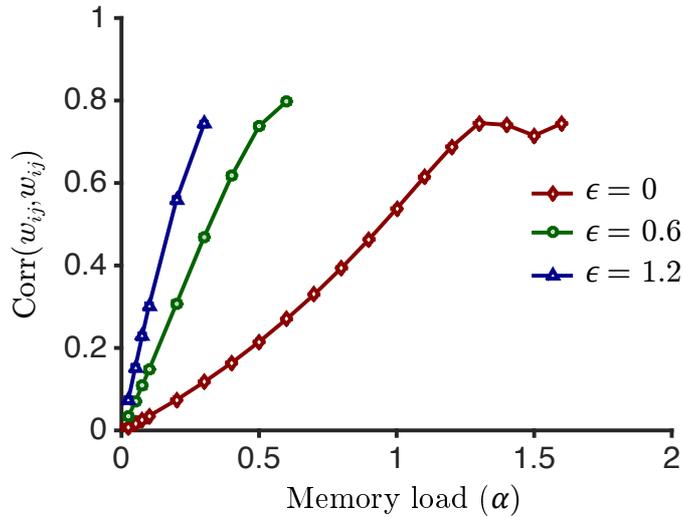}
\end{center}
\caption{\textbf{The degree of symmetry of the weight matrix.} The
  Pearson correlation coefficient between $w_{ij}$ and $w_{ji}$ is
  computed at different values of $\alpha$ for three values of
  $\epsilon$. As $\alpha$ increases the weight matrix tends to be more
  symmetric, but gets saturated for high $\alpha$. For the same values
  of $\alpha$, as the robustness increases, the correlation also
  increases, so the weight matrix becomes more symmetric. Error bars
  (across 10 runs) are smaller than the symbols.  }
\label{fig:symmetry}
\end{figure}

We have also computed the degree of symmetry of the weight matrix.
The symmetry degree is computed as the Pearson correlation coefficient
between the reciprocal weights in pairs of neurons.  We observe a
general trend towards an increasingly symmetric weight matrix as more
patterns are stored, for all values of the robustness parameter
$\epsilon$ (see Fig~\ref{fig:symmetry}).

%fig:symmetry

%f---------------------------------------------
%    DISCUSSION

\section*{Discussion}

We presented a biologically-plausible learning rule that is
characterized by three thresholds, and is able to store memory
patterns close to the maximal storage capacity in a  recurrent neural
networks without the need of an explicit ``error signal''. We
demonstrated how the learning rule can be considered a transformed
version of the PLR in the limit of a strong external field.  Our
network implements the separation between excitatory and inhibitory
neurons, with learning occurring only at excitatory-to-excitatory
synapses.  We simulated a recurrent network with $N=1001$ binary
neurons, reaching to $\alpha_c=1.6$ at zero basin size. We then used a
robustness parameter $\epsilon$ to enlarge the basin size. The
simulations showed that we are close to the theoretical capacity
across the whole investigated range of values of $\epsilon$. We expect
that as $N$ increases and the learning rate gets smaller, this
difference would go to zero.
	
Two crucial ingredients of the 3TLR are necessary: (1) strong
external inputs, (2) three learning thresholds which are set according
to the statistics of inputs to the neuron. The learning rule only uses
information that is local to a synapse and corresponding neurons. Like
classic Hebbian learning rules, our 3TLR works in an online fashion.
In addition, it can also perform as a `palimpsest'
\cite{mezard86,parisi86,amit94b}: in case the total number of patterns
exceeds the maximal capacity (at a certain basin size) the network
begins to forget patterns that are not being presented anymore.

\subsubsection*{Comparison with other learning rules}

The 3TLR can be framed in the setting of the classic Bienenstock-Cooper-Munro
(BCM) theory \cite{bienenstock82,jedlicka2002synaptic}, with additional
requirements to adapt it to the attractor network scenario. The original BCM
theory uses firing-rate units, and prescribes that synaptic modifications
should be proportional to (1) the synaptic input, and (2) a function
$\phi\left(v\right)$  of the total input $v$ (or, equivalently, of the total
output). The function $\phi\left(v\right)$ is subject to two conditions: (1)
$\phi\left(v\right)\ge0$ (or $\le0$) when $v>\theta$ (or $<\theta$,
respectively); (2) $\phi\left(0\right)=0$. The parameter $\theta$ is also
assumed to change, but on a longer time scale (such that the changes reflect
the statistics of the inputs); this (metaplastic) adaptation has the goal of
avoiding the trivial situations in which all inputs elicit indistinguishable
responses. This (loosely specified) framework ensures that, under reasonable
conditions, the resulting units become highly selective to a subset of the
inputs, and has been mainly used to model the developmental stages of primary
sensory cortex. The arising selectivity is spontaneous and completely
unsupervised: in absence of further specifications, the units become selective
to a random subset of the inputs (e.g.~depending on random initial conditions).

Our model is defined on simpler (binary) units; however, if we define
$\phi\left(v\right)=\Theta\left(v-\theta\right)\Theta\left(\theta_1-v\right)-\Theta\left(\theta-v\right)\Theta\left(v-\theta_0\right)$,
then $\phi$ behaves according to the prescriptions of the BCM theory.
Furthermore, we have essentially assumed the same slow metaplastic adaptation
mechanism of BCM, even though we have assigned this role explicitly to the
inhibitory part of the network (see Materials and Methods). On the other hand,
our model has additional requirements: (1) $\phi\left(v\right)=0$ when
$v<\theta_0$ or $v>\theta_1$, (2) plasticity occurs during presentation of
external inputs, which in turn are strong enough to drive the network towards a
desired state. The second requirement ensures that the network units become
selective to a specific subset of the inputs, as opposed to a random subset as
in the original BCM theory, and thus that they are able to collectively behave
as an attractor network. The first requirement ensures that each unit operates
close to critical capacity. Indeed, these additional requirements involve extra
parameters with respect to the BCM theory, and we implicitly assume these
parameters to also slowly adapt according to the statistics of the inputs
during network formation and development.

A variant of the BCM theory, known as ABS rule
\cite{brocher1992intracellular,artola1993long} introduced a lower threshold for
LTD, analogous to our $\theta_0$, motivated by experimental evidence; however,
a high threshold for LTP, analogous to our $\theta_1$, was not used there, or ---
to our knowledge --- in any other BCM variant. The idea of stopping plasticity
above some value of the `local field' has been introduced previously to
stabilize the learning process in feed-forward networks with discrete synapses
\cite{amit01,fusi05,brader07}. Our study goes beyond these previous works in
generalizing such a high threshold to recurrent networks, and showing that the
resulting networks achieve close to maximal capacity.

\subsubsection*{Comparison with data and experimental predictions}

In vitro experiments have characterized how synaptic plasticity
depends on voltage \cite{ngezahayo00} and firing rate
\cite{kirkwood96}, both variables that are expected to have a
monotonic relationship with the total excitatory synaptic inputs
received by a neuron. In both cases, a low value of the controlling
variable leads to no changes; intermediate values lead to depression;
and high values to potentiation. These three regimes are consistent
with the three regions for $v<\theta_1$ in Fig~\ref{fig:rule}. The
3TLR predicts that a fourth region should occur at
sufficiently high values of the voltage and/or firing rates. Most of
the studies investigating the dependence of plasticity on firing rate
or voltage have not reported a decrease in plasticity at high values
of the controlling variables, but these studies might have not
increased sufficiently such variables. To our knowledge, a single
study has found that at high rates, the plasticity vs rate curve
is a decreasing function of the input rate \cite{wang99b}.

Another test of the model consists in comparing the statistics of the
synaptic connectivity with experimental data. As it has been argued in
several recent studies
\cite{brunel04,barbour07,clopath12,chapeton12,clopath13}, networks
with plastic excitatory synapses are generically sparse close to
maximal capacity, with a connection probability that decreases with
the robustness of information storage, consistent with short range
cortical  connectivity \cite{kalisman05,song05}. Our network
is no exception, though the fraction of silent synapses that we
observe is significantly lower than in models that lack inhibition.
Furthermore, network that are close to maximal capacity tends to have
a connectivity matrix that has a significant degree of symmetry, as
illustrated by the over-representation of bidirectionally connected
pairs of neurons, and the tendency of bidirectionally connected pairs
to form stronger synapses than unidirectionally connected pairs
as observed in cortex \cite{song05,wang06},
except in barrel cortex \cite{lefort09}. Again, the 3TLR we have
proposed here reproduces this feature (Fig~\ref{fig:symmetry}),
consistent with the fact that the rule approaches the optimal
capacity.

\subsubsection*{Future directions}

Our network uses the simplest possible single neuron model
\cite{mcculloch43}. One obvious direction for future work would be to
implement the learning rule in a network of more realistic neuron
models such as firing rate models or spiking neuron models.  Another
potential direction would be to understand the biophysical mechanisms
leading to the high threshold in the 3TLR.  In any case, we believe
the results discussed here provide a significant step in the quest for
understanding how learning rules in cortical networks can optimize
information storage capacity.

\section*{Materials and Methods}

\subsection*{Simulation}

The main equations of the network, the neuron model, the learning rule, and
the criteria for stopping the learning algorithm are outlined in the Results
section, Eqs.~\ref{eq:theta}-\ref{eq:fixedpointeps}. We present here additional
details about network simulations.
	
\subsubsection*{Network setup before learning process}
	
Before applying the learning rule, we required the network to have
stable dynamics around a desired activity level $f$. A network with
only excitatory neurons is highly unstable and typically converges
towards the trivial all-off and all-on states; therefore, we
implemented a global inhibition such that the network operates around
activity level $f$. The basal inhibitory term ($H_0$) and the
inhibitory reaction term ($H_1$) are defined as:
	
	\begin{eqnarray}
		H_0 & = & (N-1)(f\bar{w}-\psi) + H^{-1}(f)\sqrt{(N-1)f\sigma_{w}}
		\label{eq:H0} \\
		H_1 & = & f\gamma\sqrt{N-1}
		\label{eq:H1}
	\end{eqnarray}
where
$H\left(x\right)=\frac{1}{2}\textrm{erfc}\left(\frac{x}{\sqrt{2}}\right)$
and $H^{-1}$ is the inverse of $H$, $\psi$ is defined as
$\theta=(N-1)\psi$; $\bar{w}$ and $\sigma_w$ are the mean and standard
deviation of the synaptic weights, respectively.  With these
definitions the network dynamics is stable in the sense that the
activity level converges to $f$ very fast, regardless of the initial
condition.
	
In Eq.~\ref{eq:H0}, we see that $H_0$ depends on the activity level
$f$ and on the standard deviation of the weights $\sigma_{w}$. In the
dense regime, $f=0.5$, we have $H^{-1}(0.5)=0$, therefore the
rightmost term of Eq.~\ref{eq:H0} vanishes, which means that in this
regime $H_0$ is independent of $\sigma_{w}$. However, in sparser
regimes, the network must be endowed with a mechanism to adjust for
the changes in standard deviation, otherwise the learning process
would bring the network out of the stable state, changing the basal
activity level. In contrast, the mean synaptic efficacy $\bar{w}$ does
not change significantly during the learning process.
 
In all our simulations, the initial values for $\{w_{ij}\}$
were sampled from a Gaussian distribution with mean and
standard deviation equal to one, after which negative values
were set to zero. This has the effect the
$\bar{w}_{ij}^\text{init}$ is slightly higher than one. We
also set $w_{ii}=0$ for all $i$.

Table~\ref{tab:param} shows the values of the parameters used
in the simulations, in the dense and sparse regimes.
\begin{table}[!ht]
\caption{\bf{Table of parameters in the simulation}}
\begin{tabular}{l l l}
\hline
\textbf{Parameter name} & \textbf{Value in dense regime} & \textbf{Value in sparse regime}\\
\hline
	$N$ & 1001 & 1001 \\
	 $\lambda=\bar{w}_{ij}^\text{init}$ & $\approx$1.08 & $\approx$1.08 \\		
	 $f$  & 0.5 & 0.2 \\
	 $\psi$ & 0.35 & 0.35\\
	 $\theta$ & 350 & 350 \\
	 $\eta$ & 0.01 [0.001 when $\epsilon=0$] & 0.01 [0.001 when $\epsilon=0$] \\
	 $\gamma$ & 6.0 & 12.0 \\
	 \# of interations (learning)  & 1000 [10000 when $\epsilon=0$] & 1000 [10000 when $\epsilon=0$]  \\
	 \# of trials in test phase  & 50 & 50 \\
\hline
\end{tabular}
\label{tab:param}
 \end{table}	

\subsubsection*{Direct comparison between the 3TLR and the PLR}

\begin{figure} 
%\internallinenumbers
\begin{center}
\includegraphics[scale=.7]{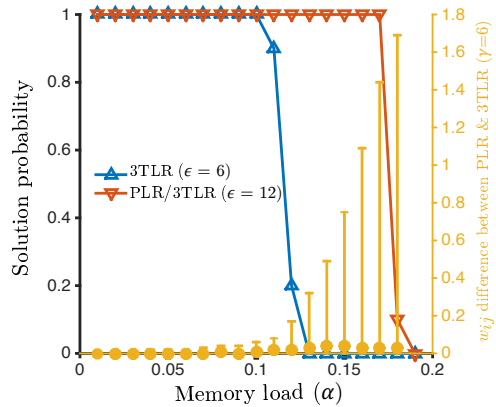}
\end{center}
\caption{\textbf{Direct comparions of the 3TLR and the PLR.}
  Success probability for the 3TLR at $\gamma=6$ (blue curve, left axis) and the
  PLR (red curve); the results for the 3TLR at $\gamma=12$ are identical to
  those of the PLR (red curve). The orange points show the absolute difference of weights  between the final values of the weights for the PLR at $\gamma=6$ and the PLR (right axis): the points show the median of the
  distribution, while the error bars span the $5$th-$95$th percentiles, showing
  that, while the distribution is concentrated at near-zero values, outliers appear
  at the critical capacity of the 3TLR algorithm. (Note that the average value of
  the weights is in all cases approximately $1.08$; also compare the discrepancies
  with the overall distribution of the weights, Fig~\ref{fig:weightdistribution}.)
  }
\label{fig:3TLR_vs_PLR}
\end{figure}

In order to determine the degree to which the 3TLR is able to mimic the
PRL, and the effect of deviations from the latter rule, we tested both rules on
the same tasks. In these simulations, every part of the simulation code was
kept identical --- including the pseudo-random numbers used to choose the
initial state and the arbitrary permutations for the update order of the units
--- except for the learning rule. We tested the network in the dense case
$f=0.5$, at $\epsilon=3$, varying the storage load $\alpha$, using $10$ samples
for each point. We compared the probability of solving the learning task and
the distribution of the discrepancies (absolute value of the differences) in
the values of the resulting synaptic weights. We tested two values of the
parameter $\gamma$, $6$ (as in Fig~\ref{fig:capacitybasin}) and $12$.  We
found that at $\gamma=12$ there was absolutely no difference between the two
rules, while at $\gamma=6$ the 3TLR performed slightly worse, and significant
deviations from the PLR started to appear close to the maximal capacity of
the 3TLR (see Fig~\ref{fig:3TLR_vs_PLR}).

\subsection*{Analytical calculation of the storage capacity at infinite $N$}

\subsubsection*{Entropy calculation}

In this section, we present the details of the calculations for the
typical storage capacity of our network in the limit of $N\to\infty$,
using the Gardner analysis \cite{gardner88,brunel04}.

The capacity is defined as the maximum value of $\alpha=p/N$ such
that a solution to Eq.~\ref{eq:fixedpointeps} can typically be found.

We can rewrite Eq.~\ref{eq:fixedpointeps} as
\begin{equation}
\forall i:\quad\prod_{\mu=1}^{\alpha N}\Theta\left(\left(2\xi_{i}^{\mu}-1\right)\left(\sum_{j=1}^{N}w_{ij}\xi_{i}^{\mu}-H_{0}-\lambda\left(\sum_{j=1}^{N}\xi_{j}^{\mu}-fN\right)-\theta\right)-f\epsilon\sqrt{N}\right)=1\label{eq:constr_1}
\end{equation}
where
\begin{eqnarray}
H_{0} & = & Nf\bar{w}-\theta+H^{-1}\left(f\right)\sigma_{w}\sqrt{fN}\\
\lambda & = & \bar{w}
\end{eqnarray}

Eq.~\ref{eq:constr_1} becomes:
\begin{equation}
\forall i:\quad\prod_{\mu=1}^{\alpha N}\Theta\left(\left(2\xi_{i}^{\mu}-1\right)\left(\sum_{j=1}^{N}\left(w_{ij}-\bar{w}\right)\xi_{i}^{\mu}-H^{-1}\left(f\right)\sigma_{w}\sqrt{fN}\right)-f\epsilon\sqrt{N}\right)=1\label{eq:constr_2}
\end{equation}

Let us now consider a single unit $i$. We write
$\sigma_{i}^{\mu}=\left(2\xi_{i}^{\mu}-1\right)$, and re-parametrize
the weights as
$W_{ij}=\frac{w_{ij}}{\bar{w}}-1\in\left[-1,\infty\right)$, and also
define
\begin{eqnarray}
T & = & H^{-1}\left(f\right)\sqrt{f}\\
K & = & \frac{\epsilon}{\bar{w}}.
\end{eqnarray}
Dropping the index $i$ and neglecting terms of order $1$, we
obtain:
\begin{equation}
\prod_{\mu=1}^{\alpha N}\Theta\left(\sigma^{\mu}\left(\sum_{j=1}^{N}W_{j}\xi_{j}^{\mu}-T\frac{\sigma_{w}}{\bar{w}}\sqrt{N}\right)-fK\sqrt{N}\right)=1\label{eq:constr_3}
\end{equation}

Our goal is to compute the quenched entropy of this problem, i.e.~the
scaled average of the logarithm of the volume of $W$ which satisfies
the above equation:
\begin{eqnarray}
S & = & \frac{1}{N}\left\langle \log V\right\rangle _{\left\{ \xi^{\mu},\sigma^{\mu}\right\} }\nonumber \\
 & = & \frac{1}{N}\left\langle \log\int\prod_{j=1}^{N}\left(dW_{j}\Theta\left(W_{j}+1\right)\right)\prod_{\mu=1}^{\alpha N}\Theta\left(\sigma^{\mu}\left(\sum_{j=1}^{N}W_{j}\xi_{j}^{\mu}-\frac{\sigma_{w}}{\bar{w}}T\sqrt{N}\right)-fK\sqrt{N}\right)\right\rangle _{\left\{ \xi^{\mu},\sigma^{\mu}\right\} }
\end{eqnarray}

The computation proceeds along the lines of \cite{gardner88,brunel04},
by using the so-called replica trick to perform the average of the
logarithm of $V$, exploiting the identity:
\begin{equation}
\left\langle \log V\right\rangle =\lim_{n\to0}\frac{\left\langle V^{n}\right\rangle -1}{n},
\end{equation}
performing the computation for integer values of $n$ and using an
analytical continuation to perform the limit $n\to0$. 
we perform the calculation using the replica-symmetric (RS) Ansatz,
which is believed to give exact results in the case of perceptron
models with continuous weights. The final expression for the entropy
depends on six order parameters; the first three are $Q$, $q$ and
$M$, whose meaning is
\begin{eqnarray*}
Q & = & \frac{1}{N}\sum_{j}\left(W_{j}\right)^{2}\\
q & = & \frac{1}{N}\sum_{j}W_{j}^{a}W_{j}^{b}\\
M & = & \frac{1}{\sqrt{N}}\sum_{j}W_{j}
\end{eqnarray*}
where we used $W^{a}$ and $W^{b}$ to denote two different \emph{replicas}
of the system, which can simply be interpreted as two independent
solutions to the constraint equation. $Q$ is called the self-overlap,
and is equal to $\left(\frac{\sigma_{w}}{\bar{w}}\right)^{2}$ in
our case, while $q$ is the mutual-overlap. The remaining order parameters
are the conjugate quantities $\hat{Q}$, $\hat{q}$ and $\hat{M}$.
The entropy expression is:
\begin{equation}
S\left(Q,q,M,\hat{Q},\hat{q},\hat{M}\right)=-\left(Q\hat{Q}-\frac{q\hat{q}}{2}\right)+\alpha\mathcal{Z}_{A}\left(Q,q,M\right)+\mathcal{Z}_{W}\left(\hat{Q},\hat{q},\hat{M}\right)
\end{equation}
where
\begin{eqnarray}
\mathcal{Z}_{A}\left(Q,q,M\right) & = & \int Du\left\langle \ln\left(H\left(\frac{K-\sigma\left(M-T\sqrt{Q}\right)+u\left(1-f\right)\sqrt{q}}{\left(1-f\right)\sqrt{Q-q}}\right)\right)\right\rangle _{\sigma}\\
\mathcal{Z}_{W}\left(\hat{Q},\hat{q},\hat{M}\right) & = & \int Du\:\ln\left(\int_{-1}^{\infty}W\, e^{-\frac{1}{2}\left(\hat{q}-2\hat{Q}\right)W^{2}+W\left(u\sqrt{\hat{q}}-\hat{M}\right)}\right).
\end{eqnarray}
We used the usual notation $Du\equiv
du\,\frac{e^{-\frac{u^{2}}{2}}}{\sqrt{2\pi}}=du\, G\left(u\right)$ to
denote Gaussian integrals, and defined
$H\left(x\right)=\int_{x}^{\infty}Du=\frac{1}{2}\textrm{erfc}\left(\frac{x}{\sqrt{2}}\right)$.
In the following, we will also use the shorthand
$\mathcal{G}\left(x\right)=\frac{G\left(x\right)}{H\left(x\right)}$.
We also used the notation $\left\langle \cdot\right\rangle _{\sigma}$
to denote the average over the output $\sigma$, i.e.~$\left\langle
\varphi\left(\sigma\right)\right\rangle
_{\sigma}=f\varphi\left(1\right)+\left(1-f\right)\varphi\left(-1\right)$
for any function $\varphi$. The value of the order parameters is found
by extremizing $S$. The notation and the following computations can be
simplified using:
\begin{eqnarray}
\Delta Q & = & Q-q\\
t_{\sigma}\left(u\right) & = & \frac{K-\sigma\left(M-T\sqrt{Q}\right)+u\left(1-f\right)\sqrt{q}}{\left(1-f\right)\sqrt{\Delta Q}}\\
\Delta\hat{Q} & = & \hat{q}-2\hat{Q}\\
\nu\left(u,W\right) & = & e^{-\frac{1}{2}\Delta\hat{Q}W^{2}+W\left(u\sqrt{\hat{q}}-\hat{M}\right)}
\end{eqnarray}

The extremization of $S$ then results in the system of equations: 
\begin{eqnarray}
\Delta\hat{Q} & = & \frac{\alpha}{\sqrt{\left(Q-\Delta Q\right)\Delta Q}}\int Du\, u\left\langle \mathcal{G}\left(t_{\sigma}\left(u\right)\right)\right\rangle _{\sigma}\label{eq:saddle_1st}\\
\hat{q} & = & \frac{\alpha}{\Delta Q}\int Du\,\left\langle \mathcal{G}\left(t_{\sigma}\left(u\right)\right)t_{\sigma}\left(u\right)\right\rangle _{\sigma}+\Delta\hat{Q}\\
0 & = & \int Du\left\langle \mathcal{G}\left(t_{\sigma}\left(u\right)\right)\sigma\right\rangle _{\sigma}\\
Q & = & \int Du\:\frac{\int_{-1}^{\infty}dW\, W^{2}\nu\left(u,W\right)}{\int_{-1}^{\infty}dW\nu\left(u,W\right)}\\
\Delta Q & = & \frac{1}{\sqrt{\hat{q}}}\int Du\: u\frac{\int_{-1}^{\infty}dW\, W\nu\left(u,W\right)}{\int_{-1}^{\infty}dW\nu\left(u,W\right)}\\
0 & = & \int Du\:\frac{\int_{-1}^{\infty}dW\, W\nu\left(u,W\right)}{\int_{-1}^{\infty}dW\nu\left(u,W\right)}\label{eq:saddle_last}
\end{eqnarray}

The integrals over $dW$ in the last three equations can be performed
explicitly, yielding: 
\begin{eqnarray}
Q & = & \frac{\hat{q}+\hat{M}^{2}+\Delta\hat{Q}}{\Delta\hat{Q}^{2}}+\frac{1}{\Delta\hat{Q}^{\frac{3}{2}}}\int Du\:\left(u\sqrt{\hat{q}}-\hat{M}-\Delta\hat{Q}\right)\mathcal{G}\left(-\frac{u\sqrt{\hat{q}}-\hat{M}+\Delta\hat{Q}}{\sqrt{\Delta\hat{Q}}}\right)\\
\Delta Q & = & \frac{1}{\Delta\hat{Q}}+\frac{1}{\sqrt{\Delta\hat{Q}\hat{q}}}\int Du\: u\mathcal{G}\left(-\frac{u\sqrt{\hat{q}}-\hat{M}+\Delta\hat{Q}}{\sqrt{\Delta\hat{Q}}}\right)\\
0 & = & -\frac{\hat{M}}{\Delta\hat{Q}}+\frac{1}{\sqrt{\Delta\hat{Q}}}\int Du\:\mathcal{G}\left(-\frac{u\sqrt{\hat{q}}-\hat{M}+\Delta\hat{Q}}{\sqrt{\Delta\hat{Q}}}\right)
\end{eqnarray}

\subsubsection*{Critical capacity}

At critical capacity, the space of the solutions shrinks to a point,
and the mutual overlap tends to become equal to the self overlap:
$q\to Q$, i.e.~$\Delta Q\to0$. In this limit, the conjugate order
parameters diverge as:
\begin{eqnarray}
\hat{q} & = & \frac{C}{\Delta Q^{2}}\label{eq:limscaling_C}\\
\Delta\hat{Q} & = & \frac{A}{\Delta Q}\label{eq:limscaling_A}\\
\hat{M} & = & \frac{B\sqrt{C}}{\Delta Q}\label{eq:limscaling_B}
\end{eqnarray}

Using these conditions, and calling $\alpha_{c}$ the critical value
of $\alpha$, the saddle point equations, \ref{eq:saddle_1st} to
\ref{eq:saddle_last}, become:
\begin{eqnarray}
Q & = & \frac{1}{A}\left(C-B\sqrt{C}\right)\label{eq:saddle_lim_1st}\\
A & = & H\left(B-\frac{A}{\sqrt{C}}\right)\\
0 & = & \frac{\sqrt{C}}{A}\left(G\left(B-\frac{A}{\sqrt{C}}\right)-BA\right)-\left(1-A\right)\label{eq:saddle_lim_3rd}\\
C & = & \alpha_{c}Q\left\langle \left(1+\tau_{\sigma}^{2}\right)H\left(\tau_{\sigma}\right)-\tau_{\sigma}G\left(\tau_{\sigma}\right)\right\rangle _{\sigma}\\
A & = & \alpha_{c}\left\langle H\left(\tau_{\sigma}\right)\right\rangle _{\sigma}\\
0 & = & \left\langle \sigma\left(G\left(\tau_{\sigma}\right)-\tau_{\sigma}H\left(\tau_{\sigma}\right)\right)\right\rangle _{\sigma}
\end{eqnarray}
where we defined
\begin{equation}
\tau_{\sigma}=\frac{\sigma\left(M-T\sqrt{Q}\right)-K}{\left(1-f\right)\sqrt{Q}}
\end{equation}

These equations can be solved numerically to find the six parameters
$\alpha_{c}$, $Q$, $A$, $B$, $C$ and $M$.

Note that in the special case $K=0$ these equations have a degenerate
solution with $Q=0$ and the same $\alpha_{c}$ as in the case of
unbounded synaptic weights (e.g.~$\alpha_{c}=2$ for $f=0.5$). This
is because in that case the original problem has the property that
scaling all weights by a factor of $x$ is equivalent to scaling the
boundary $\bar{w}$ by a factor of $x^{-1}$ (see Eq.~\ref{eq:constr_2});
therefore, the optimal strategy is to exploit this property by setting
$x\to0$, i.e.~effectively reducing the problem to the unbounded
case. Of course, this strategy can only be pursued up to the available
precision in a practical setting.

\section*{Acknowledgments}

%\nolinenumbers

\end{document}